\documentclass[aps,prl,twocolumn,superscriptaddress,floatfix]{revtex4-1}
\usepackage{graphicx,graphics}
\usepackage{dcolumn}
\usepackage{amsmath,amssymb,amsfonts}
\usepackage{latexsym,verbatim}
\usepackage{bm}
\usepackage{color}
\usepackage{ulem}
\usepackage[breaklinks=true,colorlinks,citecolor=blue,linkcolor=blue,urlcolor=blue]{hyperref}

\begin{document}
\title{Transport and optical properties of an electron gas in a Sierpinski carpet}
\author{Edo van Veen}
\affiliation{Radboud University, Institute for Molecules and Materials, NL-6525 AJ Nijmegen, The
Netherlands}
\author{Andrea Tomadin}
\email{andrea.tomadin@hotmail.com}
\affiliation{NEST, Istituto Nanoscienze-CNR and Scuola Normale Superiore, I-56126 Pisa, Italy}
\author{Mikhail I. Katsnelson}
\affiliation{Radboud University, Institute for Molecules and Materials, NL-6525 AJ Nijmegen, The
Netherlands}
\author{Shengjun Yuan}
\email{s.yuan@science.ru.nl}
\affiliation{Radboud University, Institute for Molecules and Materials, NL-6525 AJ Nijmegen, The
Netherlands}
\author{Marco Polini}
\affiliation{NEST, Istituto Nanoscienze-CNR and Scuola Normale Superiore, I-56126 Pisa, Italy}
\affiliation{Istituto Italiano di Tecnologia, Graphene Labs, Via Morego 30, I-16163 Genova, Italy}
\begin{abstract}
Recent progress in the design and fabrication of artificial two-dimensional (2D) materials 
paves the way for the experimental realization of electron systems moving on plane fractals. In this work, we present the results of computer simulations for the conductance and optical absorption spectrum of a 2D electron gas roaming on a Sierpinski carpet, i.e.~a plane fractal with Hausdorff dimension intermediate between one and two. 
We find that the conductance is sensitive to the spatial location of the leads and that it displays fractal fluctuations whose dimension is compatible with the Hausdorff dimension of the sample. Very interestingly, electrons in this fractal display a broadband optical absorption spectrum, which possesses sharp ``molecular'' peaks at low photon energies.
\end{abstract}
\maketitle

{\it Introduction.---} A variety of exerimental protocols that 
can be used to create artificial two-dimensional (2D) lattices for electrons, atoms, and photons are nowadays available.
Examples include schemes for creating artificial honeycomb lattices~\cite{polini_naturenano_2013}, 
where a wealth of interesting phenomena have been observed such as 
Mott-Hubbard split bands~\cite{singha_science_2011}, massless Dirac fermion behavior modified by 
pseudo-electric and pseudo-magnetic fields~\cite{gomes_nature_2012}, and photonic Floquet topological insulating states~\cite{rechtsman_nature_2013}. 
In the case of ultracold atomic gases loaded in honeycomb optical lattices, recent progress has even led to the experimental realization~\cite{jotzu_nature_2014} of the Haldane model~\cite{haldane_prl_1988}.

\begin{figure}[h!]
\centering
\includegraphics[width=0.72\linewidth]{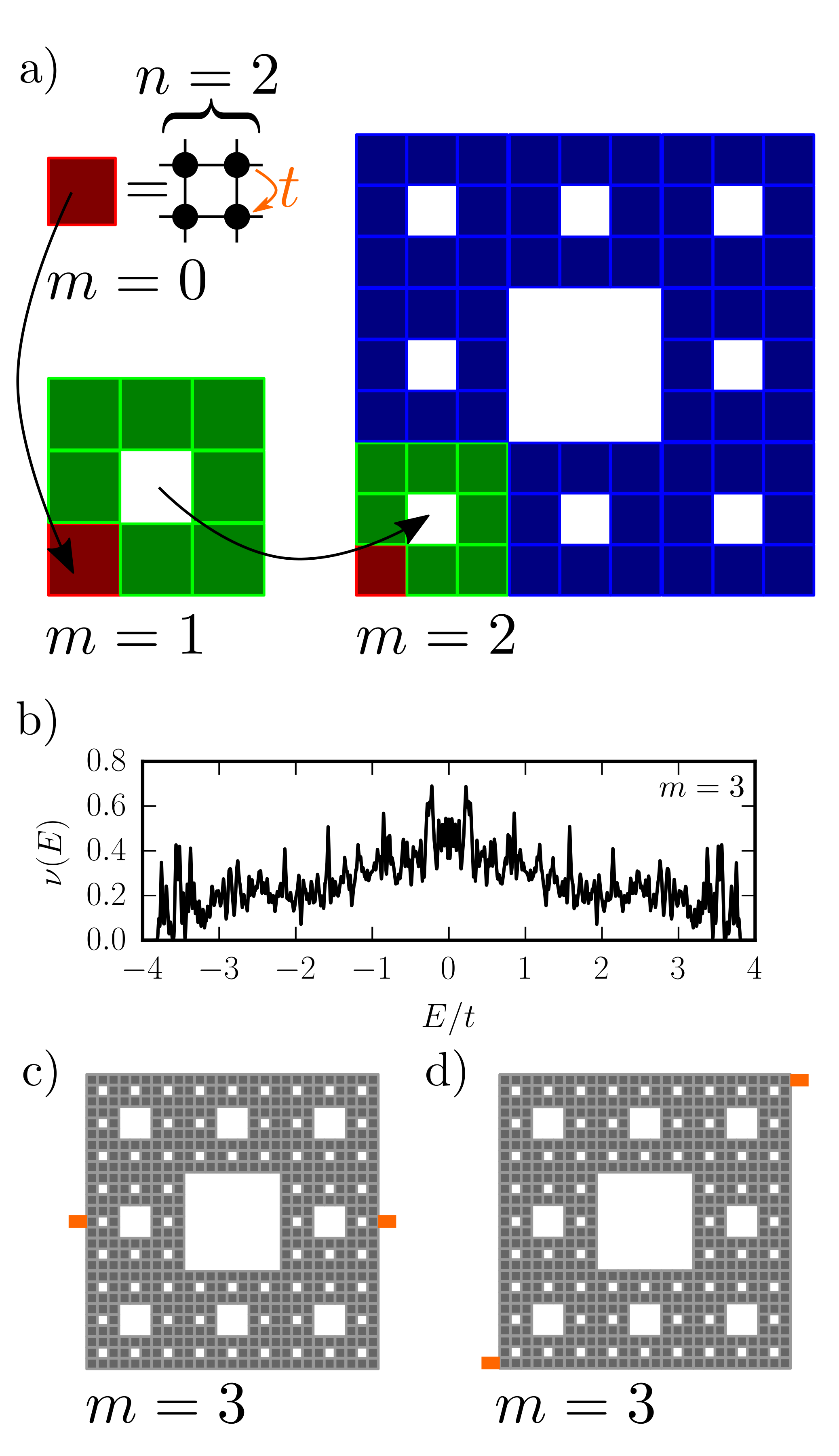}
\caption{\label{fig:one}
(Color online) Panel a) The Sierpinski carpet after $m$ iterations. The initial red unit is a $n \times n$ square lattice ($n = 2$ in this work). This unit is replicated ${\cal N} = 8$ times to obtain the green structure, which is 
${\cal L} = 3$ times longer in each of the two spatial directions and contains an inner hole with no lattice sites.
This operation is repeated to obtain the blue structure.
For a number of iterations $m \gg 1$ we obtain an approximation of the Sierpinski carpet, a plane fractal with Hausdorff dimension $d_{\rm H} = \log_{{\cal L}}{\cal N}$ ($\simeq 1.89$ in this figure). Panel b) Energy dependence of the density of states $\nu(E)$ per lattice site for a Sierpinski carpet with $m=3$.
Panel c) and d) Two examples of Sierpinski carpets with leads (orange) attached in different positions.}
\end{figure}

In the solid state, a combination of e-beam nanolithography, etching, and metallic gate deposition~\cite{gibertini_prbr_2009,desimoni_apl_2010,park_nanolett_2009,nadvornik_njp_2012,goswami_prb_2012} can in principle yield high quality two-dimensional (2D) patterns with arbitrary shape in semiconductor heterostructures (such as GaAs/AlGaAs) hosting ultra-high mobility 2D electron gases (EGs). Ultimately, these procedures yield an external potential landscape with the desired geometry that acts as a lattice of potential wells to trap electrons. The spatial resolution of these techniques can reach values of a few tens of nanometer or even below. Further improvements in spatial resolution can be obtained by bottom-up nanofabrication methods such as nanocrystal self-assembly~\cite{evers_nanolett_2013}. These approaches allow to independently control the electron density and inter-site distances and to tune the interplay between on-site and nearest-neighbor repulsive interactions and single-particle hopping, opening the way to the observation of collective phenomena and quantum phase transitions in such artificial solid-state systems~\cite{singha_science_2011}.
Synthethic solid-state quantum materials can also been created by utilizing scanning probe methods~\cite{gomes_nature_2012}. Here, suitably chosen molecules can be positioned with atomic precision on top of a substrate (such as Cu) with the aid of the tip of a scanning tunneling microscope (STM). Electrons confined in the substrate surface states and subject to the potential created by the deposited molecules can be probed via STM measurements~\cite{gomes_nature_2012}. 

These experimental achievements motivate the theoretical investigation of complex 2D structures, with the aim of discovering novel transport and optical features which could enable or improve technological applications.
In this work we present a theoretical study of the transport and optical properties of 
a 2DEG in a Sierpinski carpet (SC), which is a self-similar 2D structure~\cite{falconer_book} shown in Fig.~\ref{fig:one}. 
The self-similarity of the SC is mathematically quantified by the fact that its Hausdorff dimension~\cite{falconer_book} $d_{\rm H}$ (i.e.~a generalization of the topological dimension) is between one (a line) and two (a plane), which makes the SC a fractal~\cite{mandelbrot_book}.

While brownian motion and the heat diffusion equation 
on fractal geometries have been extensively studied in the literature~\cite{diffusionfractals,kusuoka}, the transport and optical properties of electrons roaming on such complicated geometrical structures have comparatively received less attention. More precisely, some analytical~\cite{chakrabarti_jpcm_1996,groth_prl_2008} and numerical~\cite{liu_prb_1999,lin_prb_2002,jana_physicab_2010,song_apl_2014} studies of the conductance of electrons in Sierpinski fractals have appeared in the literature. We are instead unware of studies of the optical properties of electron systems in fractals, with the exception of an experimental work~\cite{gourley_apl_1993} on a one-dimensional potential realized by growing a fractal sequence of quantum wells in (Al,Ga)As heterostructures.

In this work we show that the conductance of a 2DEG on a SC can reach the maximum value allowed by the number of open channels in the leads, depending on the lead positions and their widths, and it displays fractal fluctuations~\cite{ketzmeric_prb_1996,taylor_book_2003} as a function of energy, in the absence of a magnetic field. We show that extended states, which are responsible for large conductance values, are quite robust to elastic disorder. Finally, we calculate the real part of the optical conductivity, which turns out to be much larger than $e^2/h$ and highly structured 
up to a photon energy of the order of $t/2$, where $t$ is the hopping amplitude.

We hasten to stress that the problem at hand is very different from that of a quantum particle displaying a self-similar spectrum. Such problems are very well studied in physics, a paradigmatic example being that of the Hofstadter butterfly spectrum~\cite{hofstadter_prb_1976} displayed by an electron moving in 2D under the combined effect of a periodic potential and a perpendicular magnetic field. Finally, we are not interested in the distribution of eigenvalues and nature of the corresponding eigenstates of electrons in plane fractals, which have been studied in great detail~\cite{rammal_prl_1982,domany_prb_1983,rammal_prbr_1983,wang_prb_1995,hernando_arxiv_2015}. 
Rather, our aim is to unveil fundamental dc and ac transport characteristics, which can be measured with current technology.

{\it Model and methods.---}We describe a 2DEG in a SC by means of a single-orbital tight-binding Hamiltonian of the form
\begin{equation}\label{eq:model}
{\cal H} = -t \sum_{\langle i,j\rangle, \sigma} \left(c^\dagger_{i,\sigma}c_{j,\sigma} + {\rm H}.{\rm c}.\right)~.
\end{equation}
This describes electrons with spin $\sigma = \uparrow,\downarrow$ hopping between the nearest-neighbor sites $\langle i,j\rangle$ of a 
SC with Hausdorff dimension $d_{\rm H} = \log_{{\cal L}}{\cal N}$, as explained in Fig.~\ref{fig:one}. 
With reference to Fig.~\ref{fig:one}, each discretized SC sample is characterized by two integers, $\{n,m\}$. In this work $n=2$, while $m$ varies from $m=2$ to $m=8$. The Hamiltonian (\ref{eq:model}) is particle-hole symmetric and the spectrum of eigenvalues extends from $-4t$ to $4t$ for a bandwidth equal to $8t$.

The hopping parameter $t$ is used below as unit of energy. 
Nanopattering a SC on the surface of semiconductor hosting a high mobility 2DEG is expected~\cite{singha_science_2011,gibertini_prbr_2009,desimoni_apl_2010} to yield 
$t$ of the order of a few meV, while protocols based on STM manipulation~\cite{gomes_nature_2012} are expected to yield $t \sim 100~{\rm meV}$. For the sake of simplicity, we ignore magnetic fields and electron-electron interactions, which are expected to lead to very interesting quantitative and qualitative effects that will be the subject of future works.

The conductance and wavefunction calculations are performed by using \textsc{kwant}~\cite{groth_2014}. In this toolkit, wavefunction matching is implemented to compute the wavefunctions in the scattering region and the scattering matrix $S_{nm}$ for an incoming propagating mode $n$ and an outgoing mode $m$. The conductance between the left lead~L 
and the right lead~R is given by the Landauer formula
\begin{equation}\label{eq:landauer}
G = \frac{2e^2}{h} \sum_{n \in {\rm L}, m\in {\rm R}} |S_{nm}|^2~.
\end{equation}

The density of states (DOS) $\nu(E)$ and optical conductivity $\sigma(\omega)$ are calculated by using the tight-binding propagation method (TBPM)~\cite{yuan_prb_2010,yuan_prb_2011}. Since TBPM 
does not involve diagonalization of matrices, both CPU time and memory cost grown linearly with the sample size, allowing calculations with up to $10^{10}$ sites.

In TBPM, the DOS is obtained by the following Fourier transform~\cite{hams_prb_2000,yuan_prb_2010,yuan_prb_2011}
\begin{equation}\label{eq:DOS}
\nu (E)=\frac{1}{2\pi }\int_{-\infty }^{\infty }\exp{(iE\tau/\hbar)}\left\langle \varphi |\varphi (\tau)\right\rangle d\tau~,  
\end{equation}
where the initial state $\left\vert \varphi \right\rangle = \sum_i a_i\vert i\rangle$ is a superposition of localized orbitals $\vert i\rangle$ with complex random coefficients $a_i$ and the wave propagation $\left\vert \varphi (\tau)\right\rangle \equiv \exp{(- i {\cal H}\tau/\hbar)}\left\vert \varphi \right\rangle $ is performed numerically by using the Chebyshev polynomial algorithm. 
Our results for the DOS of a $\{2,3\}$ SC are reported in Fig.~\ref{fig:one}b). We have checked that the function $\nu(E)$ rapidly converges for a fixed value of $n$ and increasing $m$: for example, results for the $\{2,7\}$ and $\{2,8\}$ SCs (not shown) are nearly indistinguishable over the entire bandwidth.

{\it Quantum transport and fractal conductance fluctuations.---}We have calculated the energy dependence of the two-terminal conductance $G(E)$ of the tight-binding model (\ref{eq:model}), for two different lead configurations, as in Figs.~\ref{fig:one}c) and~d). 

\begin{figure}[t]
\includegraphics[width=\linewidth]{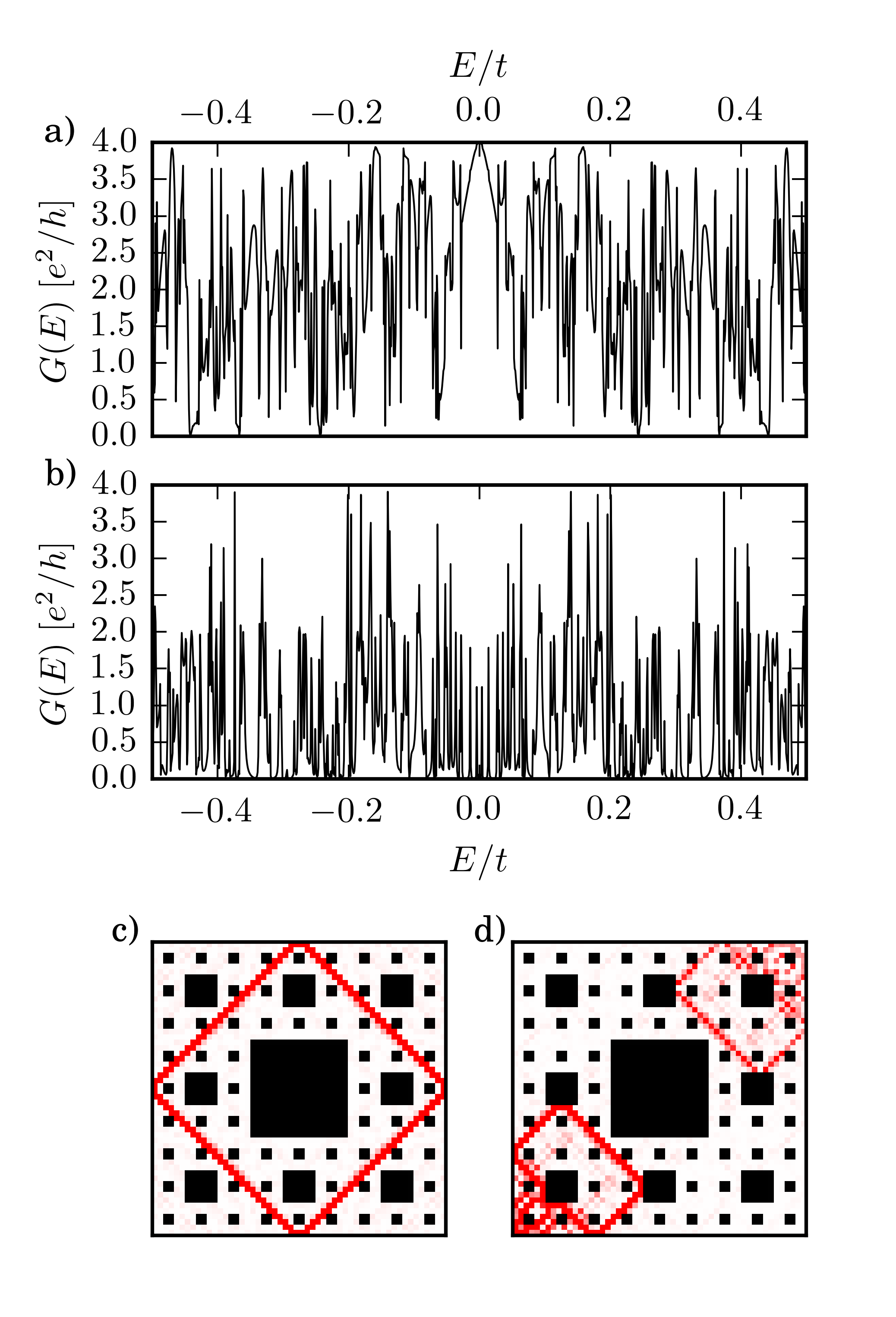}
\caption{\label{fig:two}
(Color online) Energy dependence of the conductance $G(E)$ (in units of $e^2/h$) 
of a $\{2,3\}$ Sierpinski carpet.
Data in panel~a) [panel~b)] refer to the sample with leads positioned as in Fig.~\ref{fig:one}c) [Fig.~\ref{fig:one}d)].
Panels c) and d) show the square modulus $|\psi({\bm r}_i)|^{2}$ of the electron wave function corresponding to $E = 0$, with leads positioned as in panels a) and b), respectively.
The colorscale varies from white ($|\psi({\bm r}_i)|^{2} = 0$) to red (maximum). Black squares denote holes in the Sierpinski carpet. The wave function in panel c) represents the state responsible for $G(0)=4e^2/h$ in panel a). Note the extreme geometrical simplicity of this state, which connects the two leads while avoiding the holes in the Sierpinski carpet. On the contrary, the wave function in panel d) represents an insulating state, where the electrons are totally back-scattered from the holes in the Sierpinski carpet.}
\end{figure}

A summary of our main results for a $\{2,3\}$ SC is reported in Fig.~\ref{fig:two}. In panel a) we clearly see that the two-terminal conductance $G(E)$ is equal to $4e^2/h$ for $E=0$, where a conductive extended state is present~\cite{chakrabarti_jpcm_1996}. This is because for the two leads positioned as in Fig.~\ref{fig:one}c), electrons of a given spin injected on the left side of the sample can reach the right side by following two equivalent paths, Fig.~\ref{fig:two}c), each carrying a conductance quantum, without being backscattered by the inner holes of the SC. On the other hand, as we can see from panel b), the sample can be insulating (i.e.~$G=0$), at the same energy when probed with the leads positioned as in Fig.~\ref{fig:one}d).

\begin{figure}[t]
\includegraphics[width=\linewidth]{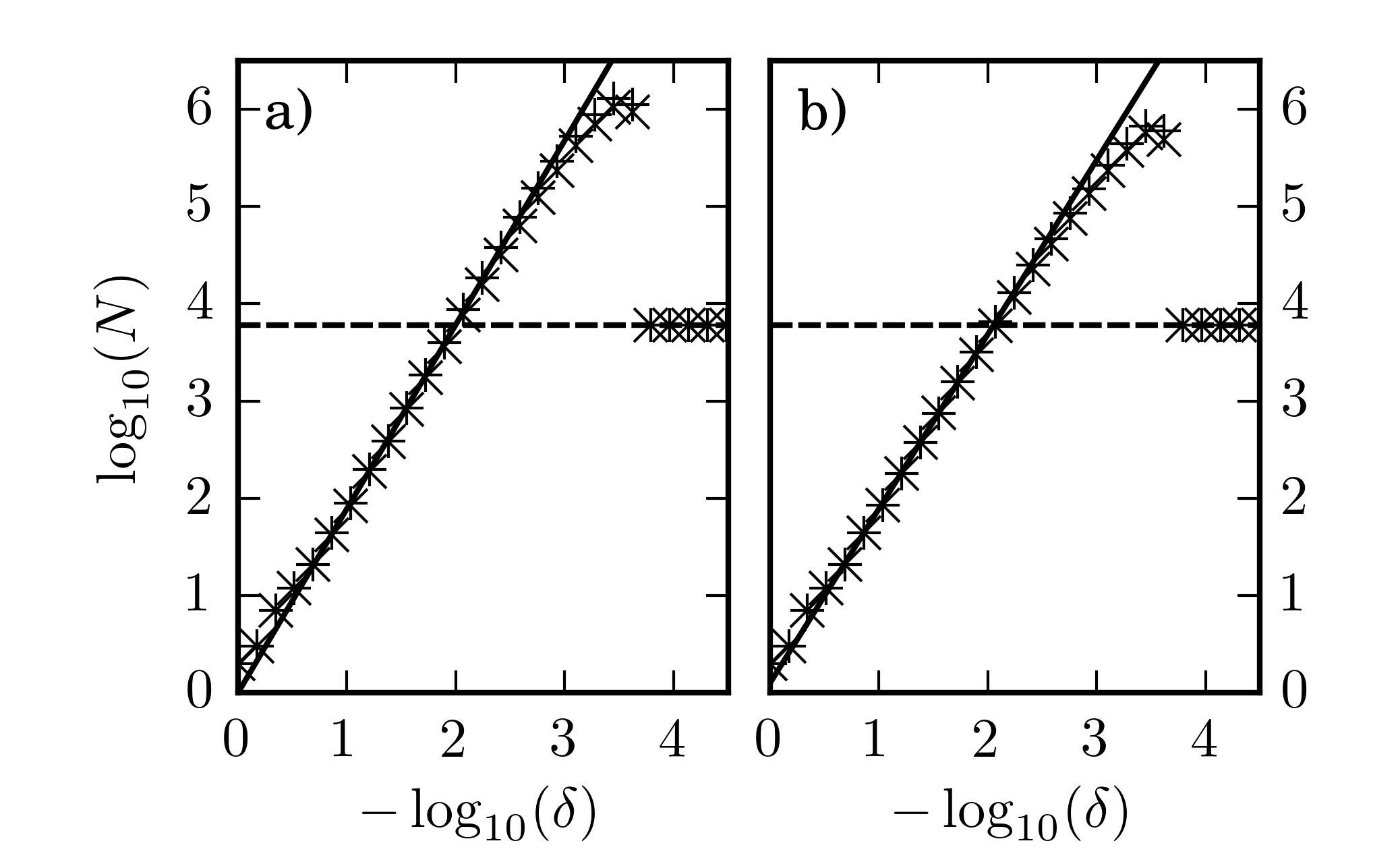}
\caption{\label{fig:three}
Box-counting analysis of the conductance fluctuations in the Sierpinski carpet.
Data in panel a) have been obtained for a Sierpinski carpet with ${\cal N} = 8$ and ${\cal L}=3$, as in Fig.~\ref{fig:one}a).
Data in panel b) have been obtained for a Sierpinski carpet with ${\cal N} = 12$ and ${\cal L}=4$.
In each panel, $+$ and $\times$ symbols represent the results of the box-counting algorithm applied to samples 
with the leads positioned as in Fig.~\ref{fig:one}c) and d), respectively.
The horizontal dashed lines represent the constant value $N = N_{\rm s}$, with $N_{\rm s} = 6000$. 
In each panel, the slope $d$ of the solid lines has been set equal to the Hausdorff dimension $d_{\rm H}$ of the corresponding Sierpinski carpet, i.e.~$d = 1.89$ in panel a) and $d = 1.79$ in panel b).}
\end{figure}

Data in Fig.~\ref{fig:two} display also wild fluctuations. Such conductance fluctuations (CFs) can be quantitied by 
using a box-counting (BC) algorithm~\cite{guarneri_pre_2001}. This counts the number $N$ of squares of size $\delta$, which is necessary to continuously cover the graph of $G(E)$ (in units of $e^2/h$) rescaled to a unit square.
In general, points in the plane $(\log{N},-\log{\delta})$ are expected to fall in {\it three} distinct regions.
For large values of $\delta$, the squares are too large to distinguish the features of the graph and $N$ grows slowly as $\delta$ decreases. For very small values of $\delta$, the squares are so small that they resolve the single points in the set of data belonging to the CF graph: in this case $N$ is expected to saturate to the number $N_{\rm s}$ of points in the energy mesh where $G(E)$ is evaluated. Finally, there is an intermediate region (usually called ``scaling region'') where scaling is linear in the log-log plane, i.e.~where $N \sim \delta^{-d}$. The slope $d$ in the scaling region is the BC estimate of the Hausdorff dimension of the CFs.

In Fig.~\ref{fig:three}, we show the results of the BC algorithm for the CFs of {\it two} SC samples with (slightly) different Hausdorff dimensions, obtained by changing ${\cal N}$ and ${\cal L}$ in the iterative geometrical construction illustrated in Fig.~\ref{fig:one}a). Results do not depend on the spatial arrangement of the leads. The analyzed CFs clearly show fractal behavior over a scaling region of more than two orders of magnitude. 
Fractal CFs emerge~\cite{ketzmeric_prb_1996}  in the problem of phase-coherent ballistic transport and in the presence to strong quantum localization~\cite{guarneri_pre_2001}. Indeed, in the SC transport problem, almost all wavefunctions are localized~\cite{domany_prb_1983} [see e.g.~Fig.~\ref{fig:two}d)] and do not contribute to transport (even in the absence of elastic disorder) due to scattering of electrons against the inner holes.

Finally, we note that the value of $d$ extracted from our numerical analysis is in excellent agreement with the Hausdorff 
dimension $d_{\rm H}$ of the corresponding SC sample~\cite{footnote_dimension}. It is remarkable that the analysis of CFs carries information on the sample geometry, down to very small length scales.

\begin{figure}[t]
\includegraphics[width=\linewidth]{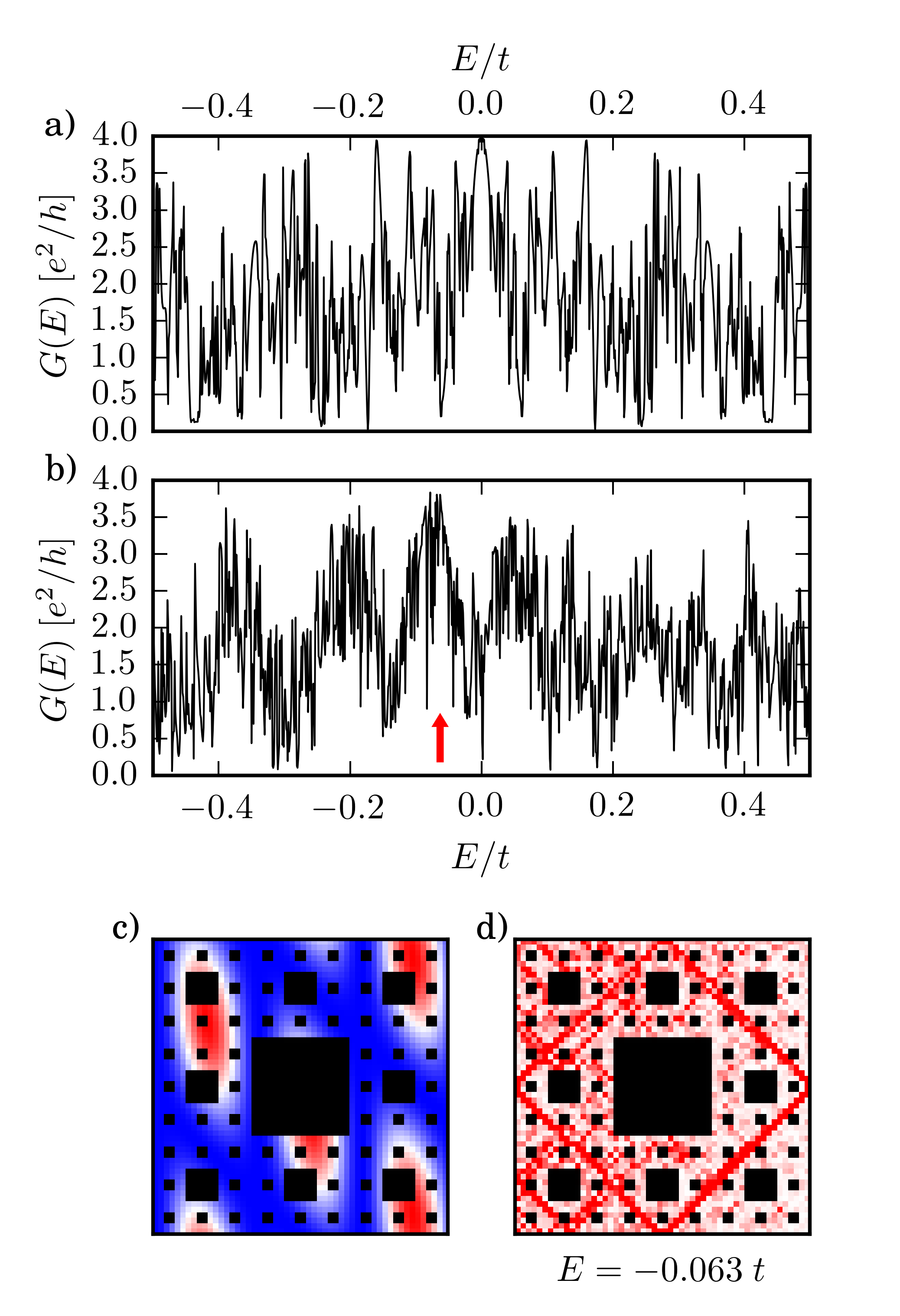}
\caption{\label{fig:four}
(Color online)
Quantum transport in a disordered Sierpinski carpet, with leads positioned as in Fig.~\ref{fig:one}c). 
Panels a) and b) show the energy dependence of the conductance $G(E)$ (in units of $e^2/h$) 
in the presence of localized and smooth elastic disorder, respectively.
Data shown in panel a) have been obtained by creating a single-site vacancy along the path of the conductive state shown in Fig.~\ref{fig:two}c), on the site of the Sierpinski carpet with spatial coordinates ${\bm r}_i = (10,18)$.
Data shown in panel b) have been obtained by adding to (\ref{eq:model}) 
a smooth disorder potential $\delta \mu({\bm r}_i)$ described by Hamiltonian (\ref{eq:disorder}).
The potential is shown in c), where the colorscale varies from $-0.1t$ (blue) to $+0.1t$ (red).
The conductive eigenstate at $E = -0.063 t$ [indicated by a red arrow in b)] is shown in d). }
\end{figure}

Fig.~\ref{fig:four} shows that the conductance of the 2DEG in a SC is robust with respect to both localized [panel a)] and smooth [panel b)] elastic disorder. 
The impact of a single-site vacancy---located on the highly-conductive path through the bulk of the SC displayed in Fig.~\ref{fig:two}c)---is shown in Fig.~\ref{fig:four}a). We see that, despite such a strong, localized disorder source, $G(E)$ still reaches its maximum value $G(E)=4e^2/h$ at $E =0$. The impact of a smooth disorder potential, 
\begin{equation}\label{eq:disorder}
V = \sum_{i} \delta \mu({\bm r}_i) c^\dagger_i c_i~,
\end{equation}
which varies on an energy scale equal to $20\%$ of the hopping amplitude, is shown in Fig.~\ref{fig:four}b). The spatial variations of $\delta \mu({\bm r}_i)/t$ are reported in Fig.~\ref{fig:four}c), while the fate of the highly-conductive bulk path shown in Fig.~\ref{fig:two}c) is illustrated in Fig.~\ref{fig:four}d).

\begin{figure}
\includegraphics[width=\linewidth]{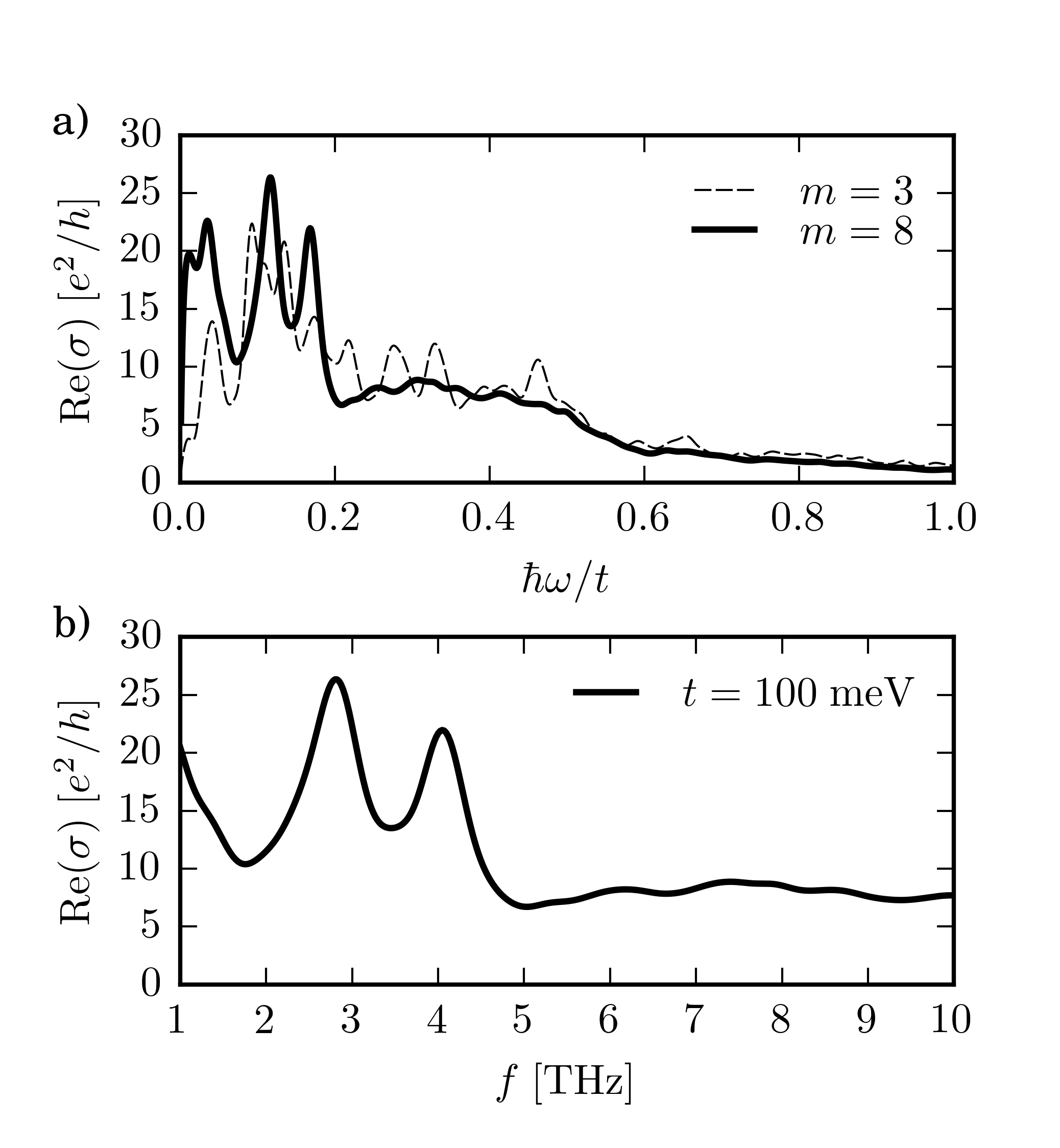}
\caption{\label{fig:five} Panel a) The real part of the optical conductivity $\sigma(\omega)$ (in units of $e^2/h$) of a 2DEG in a Sierpinki carpet is plotted as a function of photon energy $\hbar\omega$ (in units of $t$), for a fixed value of temperature $k_{\rm B} T = 0.0034 t$ (this choice implies $T = 4~{\rm K}$ for $t =100~{\rm meV}$). 
Data in this panel refer to two different values of $m$, $m=3$ (thin dashed line) and $m=8$ (thick solid line). Panel b) ${\rm Re}[\sigma(\omega)]$ (in units of $e^2/h$) 
for $m=8$,  $t =100~{\rm meV}$, and $T=4~{\rm K}$.}
\end{figure}

{\it Optical conductivity.---}We have calculated the optical conductivity from the Kubo formula for the current response function to a uniform time-dependent vector potential. In the TBPM, this is calculated  (omitting the Drude contribution at $\omega =0$) from~\cite{yuan_prb_2010,yuan_prb_2011}:
\begin{eqnarray}
\mathrm{Re}[\sigma (\omega )] &=&\frac{1}{S}\lim_{\epsilon \rightarrow 0^{+}}\frac{
e^{-\beta \hbar \omega }-1}{\omega}\int_{0}^{\infty }e^{-\hbar \epsilon
\tau}\sin{(\hbar \omega \tau)}  \notag  \label{eq:optical} \\
&&\times 2~\text{Im}\left\langle \varphi |f({\cal H}) J(
\tau) [1-f({\cal H})] J|\varphi \right\rangle
d\tau~,
\end{eqnarray}
where $S$ is the sample area, $\beta =1/(k_{\rm B}T)$ is the inverse temperature, $f({\cal H}) =1/\left[ e^{\beta \hbar({\cal H} -\mu) }+1\right] $ is the Fermi-Dirac distribution operator, and $J(\tau) =e^{i{\cal H}\tau/\hbar}Je^{-i{\cal H}\tau/\hbar}$ is 
the current operator in the Heisenberg picture of the time evolution.

Fig.~\ref{fig:five}a) shows that the dependence of the absorption spectrum ${\rm Re}[\sigma(\omega)]$ (in units of $e^2/h$) on the photon energy $\hbar\omega$ (in units of $t$). This quantity is evaluated at a finite but small temperature $k_{\rm B} T\ll t$ (specifically, $k_{\rm B} T= 0.0034 t$) and for two SC samples, i.e.~$\{2,3\}$ and $\{2,8\}$. By looking at the difference between ${\rm Re}[\sigma(\omega)]$ at $m=7$ (not shown) and $m=8$ (and fixed temperature), we conclude that the data shown for $m=8$ represent the $m\to \infty$ result. We note that the optical absorption spectrum features three distinct behaviors. For photon energies $\hbar \omega \lesssim 0.2 t$, absoption is very large, with ${\rm Re}[\sigma(\omega)] \gg e^2/h$, peaking at $\approx 25 e^2/h$. The peak value is quite sensitive to the value of $n$ in a $\{n,m\}$ SC: for example, 
this peak reaches $\approx 40 e^2/h$ in a $\{4,8\}$ SC. For photon energies $0.2t \lesssim \hbar \omega \lesssim 0.5t$, ${\rm Re}[\sigma(\omega)]$ features a plateau at a value which is intermediate between $5 e^2/h$ and $10 e^2/h$. Finally, an exponential tail 
${\rm Re}(\sigma) \propto e^{-\alpha (\hbar\omega/t)}$ with $\alpha \approx 1.5$ kicks in for $\hbar \omega > 0.5 t$.

For a choice of the hopping amplitude $t = 100~{\rm meV}$ (which is the expected value for a SC realized by STM manipulation~\cite{gomes_nature_2012}), the absorption of the SC vanishes in the visible range of the electromagnetic spectrum and it is of the order of $e^2/h$ in the infrared. For photon frequencies spanning the range 
$300~{\rm GHz} \leq f \leq 5~{\rm THz}$ the real part of the optical conductivity is very large, much larger than $e^2/h$, and displays a sequence of molecular resonances (whose smoothness is controlled by temperature).

{\it Summary.---}In this work we have carried out extensive computer simulations of the conductance and optical absorption spectrum of a two-dimensional electron gas roaming on a prototypical plane fractal, the Sierpinski carpet. We have demonstrated that the conductance is sensitive to the spatial location of the leads and that it displays fractal fluctuations whose dimension is compatible with the Hausdorff dimension of the sample. Due to the presence of a massive number of localized eigenstates, the optical absorption spectrum turns out to display sharp molecular peaks at low photon energies, which may pave the way for applications in sensing and plasmonics.

\acknowledgments
This work was supported by the European Research Council Advanced Grant program (contract 338957) (S.Y. and M.I.K.) and by the Italian Ministry of Education, University, and Research (MIUR) through the program ``Progetti Premiali 2012" - Project ABNANOTECH (A.T. and M.P.). Support by the Netherlands National Computing Facilities foundation (NCF) is gratefully acknowledged. We gratefully acknowledge Luigi Ambrosio and Fabio Taddei for useful discussions, and Carlo Beenakker for very useful correspondence.


\begin{thebibliography}{77}
\bibitem{footnote_dimension}
	The Hausdorff dimensions of the SCs used to produce the results in Figs.~\ref{fig:three}a) and b) 
	differ by only $5\%$. To further corroborate the conjecture $d=d_{\rm H}$ one needs to carry out a more detailed data
	analysis in which SCs with larger values of ${\cal N}$ are used. This 
	lowers the fractal dimension and can help finding deviations from the conjecture $d=d_{\rm H}$.
	Such a detailed numerical investigation requires the simulation of large samples, which is quite demanding, 
	and lies outside the scope of the present work.
\end{thebibliography}
\end{document}